\documentclass[aip,apl,reprint,groupedaddress,preprintnumber,showpacs]{revtex4-1}
\usepackage{amsmath}
\usepackage{amssymb}
\usepackage{graphicx}
\usepackage{amssymb}
\usepackage{graphics}
\usepackage{epsfig}
\usepackage{CJK}
\usepackage{color}
\usepackage{graphicx}
\usepackage{epstopdf}
\usepackage{multirow}
\usepackage{tabularx}
\begin{document}

\title{Thermal transport and mixed valence in ZrTe$_3$ doped with Hf and Se}
\author{Yu Liu,$^{1,\S,\dag}$ Zhixiang Hu,$^{1,2}$ Xiao Tong,$^{3}$ Denis Leshchev,$^{4}$ Xiangde Zhu,$^{1,*}$ Hechang Lei,$^{1,\|}$ Eli Stavitski$^{4}$, Klaus Attenkofer,$^{4,\P}$  and C. Petrovic$^{1,2,\ddag}$}
\affiliation{$^{1}$Condensed Matter Physics and Materials Science Department, Brookhaven National Laboratory, Upton, New York 11973, USA\\
$^{2}$Department of Material Science and Chemical Engineering, Stony Brook University, Stony Brook, New York 11790, USA\\
$^{3}$Center of Functional Nanomaterials, Brookhaven National Laboratory, Upton, New York 11973, USA\\
$^{4}$National Synchrotron Light Source II, Brookhaven National Laboratory, Upton, NY 11973, USA\\
}
\date{\today}

\begin{abstract}
Two-dimensional transition metal trichalcogenides (TMTC's) feature covalently bonded metal-chalcogen layers separated by the van der Waals (vdW) gap. Similar to transition metal dichalcogenides (TMDCs), TMTCs often host charge density waves (CDWs) and superconductivity but unlike TMDCs atomic chains in the crystal structure give rise to quasi one-dimensional (quasi 1D) conduction. ZrTe$_3$ features CDW below $T_{\textrm{CDW}}$ = 63 K and filamentary superconductivity below 2 K that can be enhanced by pressure or chemical substitution. Here we report the presence of mixed valent Zr$^{2+}$ and Zr$^{4+}$ atoms in ZrTe$_3$ crystals that is reduced by doping in ZrTe$_{3-x}$Se$_x$ and Zr$_{1-y}$Hf$_y$Te$_3$. Superconductivity is enhanced via disorder in Te2-Te3 atomic chains that are associated with CDW formation. Hf substitution on Zr atomic site enhances $T_{\textrm{CDW}}$ due to unperturbed Te2-Te3 chain periodicity and enhanced electron-phonon coupling. Weak electronic correlations in  ZrTe$_{3-x}$Se$_x$  are likely governed by the lattice contraction effects.
\end{abstract}
\maketitle

Interplay between the charge density wave (CDW) and conventional superconductivity (SC), both Fermi surface instabilities and low-temperature collective orders caused by strong electron-phonon coupling, has been a subject of extensive investigations over past decades.\cite{Gurner} CDW instability commonly arises in a metallic one-dimensional (1D) chain at zero temperature due to Fermi surface nesting, i.e. energetically favorable lattice reconstruction under electronic perturbation with momentum space peridicity of $q$ = 2$k_\textrm{F}$ where $k_\textrm{F}$ is the Fermi wavevector; in the case of a single half-filled band this leads to metal-insulator transition.\cite{Peierls} In 1D metals with Kohn anomaly CDW is typically found below temperature $T_{\textrm{CDW}}$, developing a BCS-type energy gap.\cite{Rice} Over the course of years, it has been recognized that CDW may also arise due to strong enhancement of electron-phonon coupling at some wavevector unrelated to nesting condition or to electron-electron interaction in materials with strong Coulomb energy.\cite{ZhuXPNAS,ZhuXAPX}

Layered ZrTe$_3$, an interesting metallic member of MX$_3$ (M = Hf, Zr; X = S, Se, Te) TMTCs, has been attracting extensive attention. It features low-dimensional atomic arrangement in its unit cell of $P2_1/m$ symmetry [Fig. 1(a)].\cite{Bra,Fur,Wieting,RandleM} ZrTe$_3$ undergoes a nesting-type CDW transition that opens only a partial gap at the Fermi surface below $T_{\textrm{CDW}}$ $\sim$ 63 K since multiple bands cross the Fermi surface. Its crystal structure is quasi two-dimensional (2D) with vdW gap, but it hosts two quasi 1D trigonal prismatic ZrTe$_{6}$ chains with inversion symmetry that propagate along the $b$-axis; in addition there are Te2-Te3 chains along the $a$-axis [Fig. 1(a)]. CDW originates from nesting in an electron pocket with highly directional Te 5$p_x$ orbital character along the chains whereas other parts of the Fermi surface are unaffected.\cite{Eaglesham,Takahash,Nakajima,Yamaya,Tsu,XD2,Felser} In contrast to NbSe$_3$, CDW-induced resistivity anomaly is observed in electrical resistivity for the current path along the $a$-axis but is absent for the current path along the $b$-axis due to the nesting wavevector $\overrightarrow{q}$ $\equiv (\frac{1}{14},0,\frac{1}{3})$.\cite{Eaglesham,Hod,Takahashi2} Band structure calculation and angular resolved photoemission (ARPES) measurements revealed that the Fermi surface (FS) consists of a three-dimensional (3D) FS sheet at the Brillouin zone (BZ) center and quasi-1D FS sheets parallel to the inclination of the BZ boundary.\cite{Stowe,Star,Hoesch,Lyu} The Kohn anomaly associated with a soft phonon mode and CDW fluctuations have been identified.\cite{Per,Yokoya,Chinotti,Hoesch1} Whereas strong electron-phonon coupling is important for CDW formation in ZrTe$_3$, structural changes are detected at the onset of pressure-induced superconductivity.\cite{Hu,Gleason}

Below $T_{\textrm{CDW}}$, ZrTe$_3$ shows a filamentary-to-bulk SC at $T_\textrm{c}$ $\sim$ 2 K with local pair fluctuations; SC first condenses into filaments along the $a$-axis, and then becomes phase coherent below 2 K. Bulk SC with an enhanced $T_\textrm{c}$ is observed by applying pressure, intercalation, substitution, and disorder, with suppression of CDW order.\cite{Yomo,Hoesch2,Gu,HC,XD1,Yadav,Zhu,Cui,Ganose,Yan,Yue} Pressure-induced re-entrant SC in ZrTe$_3$ implies the possible unconventional Cooper pairing mechanism, \cite{Yomo} yet the ultra-low-temperature thermal conductivity indicates multiple nodeless gaps in ZrTe$_{3-x}$Se$_x$.\cite{Cui} ZrTe$_{3-x}$Se$_x$ displays SC with the $T_c$ up to 4.4 K for $x$ $\sim$ 0.04, where the CDW-related modes in Raman spectra is observable while the long-range CDW order vanishes.\cite{XD2} In contrast to ZrTe$_3$, the isostructural HfTe$_3$ undergoes a CDW transition at $T_{\textrm{CDW}}$ $\sim$ 93 K without the appearance of SC down to 50 mK at ambient pressure; the SC pairing starts to occur only within the 1D Te2-Te3 chain but no phase coherence between the SC chains can be realized under pressure.\cite{Saleem,JingLi,JGC}

In superconducting ZrTe$_{2.96}$Se$_{0.04}$ CDW fluctuation-induced electronic correlations were proposed since heavy-fermion-like mass enhancement of mass tensor anisotropy has been detected.\cite{XD2} In addition, mixed valence of Zr in ZrTe$_3$ nanoribons was observed.\cite{YuX} Valence segregation is associated with superconductivity in transition metal oxides with both weak and strong electronic correlations and has been discussed in connection with superconducting mechanism and electron-phonon coupling.\cite{GoodenoughJB,VarmaCM,RiceTM,KimM} Thermal transport is an efficient method to characterize the nature and sign of carries as well as the correlation strength in superconductors whereas X-ray photoemission spectroscopy (XPS) and Raman measurements are good probes of the valence state and phonon vibrations in transition metal compounds.\cite{Kefeng,Behnia,Pourret,DemeterM}

Here we examine electronic correlation strength and Zr valence in superconducting ZrTe$_3$ single crystals doped with Se on Te and contrast this with ZrTe$_3$ doped with Hf on Zr atomic site when the electrical and thermal current flow is restricted along the ZrTe$_6$ chains, i. e. the $b$-axis. We observe decrease of Zr-Te bond lengths in both Hf- and Se-doped crystals, consistent with smaller unit cells of HfTe$_3$ and ZrSe$_3$. Thermal transport and Raman measurements show increase in $T_{\textrm{CDW}}$ with Hf substitution and a rapid suppression of $T_{\textrm{CDW}}$ with Se doping. Significant mixed-valent disproportion in ZrTe$_3$ is reduced in bulk superconducting ZrTe$_{2.96}$Se$_{0.04}$ as well as in non-superconducting Zr$_{0.95}$Hf$_{0.05}$Te$_3$. Tendency towards equivalent metal valence with completely suppressed CDW in ZrTe$_{2.96}$Se$_{0.04}$ is similar to superconducting Ba$_{1-x}$K$_x$SbO$_3$.\cite{KimM}

Single crystals of ZrTe$_{3-x}$Se$_x$ ($x$ = 0, 0.01, 0.04) and Zr$_{1-y}$Hf$_y$Te$_3$ ($y$ = 0.01 and 0.05) were fabricated by the chemical vapor transport method.\cite{XD2} Mixture of pure elements Hf, Zr, Te, and Se powder were sealed with $\sim$ 5 mg cm$^{-3}$ iodine as the transport agent in an evacuated quartz tube. The furnace gradient was kept between 1023 and 923 K after heating at 973 K for two days. The actual elemental ratio was checked by using energy-dispersive x-ray spectroscopy in a JEOL JSM-6500 scanning electron microscope (SEM). The X-ray absorption near edge structure (XANES) and extended X-ray absorption fine structure (EXAFS) spectra were measured at 8-ID beamline of the National Synchrotron Light Source II (NSLS II) at Brookhaven National Laboratory (BNL) in the fluorescence mode, and processed using the Athena software package. The extracted EXAFS signal $\chi(k)$ was weighed by $k^3$ to emphasize the high-energy oscillation and then Fourier-transformed to analyze the data in $R$ space. Thermopower and electrical resistivity were measured in a Quantum Design PPMS-9 with standard four-probe technique with thermal gradient and electrical current flow directed along the $b$-axis. The sample dimensions were measured by an optical microscope Nikon SMZ-800 with 10 $\mu$m resolution. XPS measurements were carried out in an ultrahigh-vacuum (UHV) system with 3$\times$10$^{-10}$ Torr base pressure, equipped with a SPECS Phoibos 100 spectrometer and a non-monochromatized Al-K$_{\alpha}$ X-ray source ($h$$\nu$ = 1486.6 eV). XPS peak positions were calibrated using metallic  Te 3$d$$_{5/2}$ at 573.0 eV. Single selected point unpolarized Raman spectrum experiment was performed using WITec confocal Raman microscope alpha 300 equipped with an solid-state laser ($\lambda$ = 532 nm), an electron multiplying CCD detector and an 100$\times$/0.9NA objective lens. Raman scattered light was focused onto a multi-mode fiber and monochromator with a 1800 line/mm grating. In XPS and Raman measurements samples were sputtered in UHV by 2$\times$10$^{-5}$ Torr of Ar$^+$ ions with kinetic energy of 2500 eV for 60 min in order to remove surface oxygen contamination. The Raman spectra  were measured  right after the samples were taken out from UHV chamber. The Raman shows no difference between Ar sputtered and freshly exfoliated samples.

\begin{figure}
\centerline{\includegraphics[scale=1]{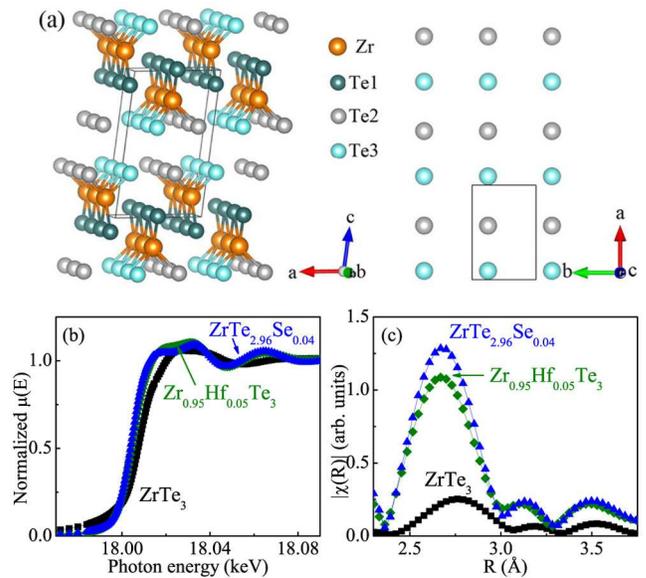}}
\caption{(Color online) Crystal structure of ZrTe$_3$ with quasi-one-dimensional (1D) ZrTe$_6$ chains along the $b$-axis and 1D Te2-Te3 chains along the $a$-axis. (b) Normalized Zr K-edge XANES and (c) Fourier transform magnitudes of EXAFS data of indicated samples. The curves in (c) represent raw experimental data without correcting for the phase shifts.}
\end{figure}

\begin{figure*}
\centerline{\includegraphics[scale=1]{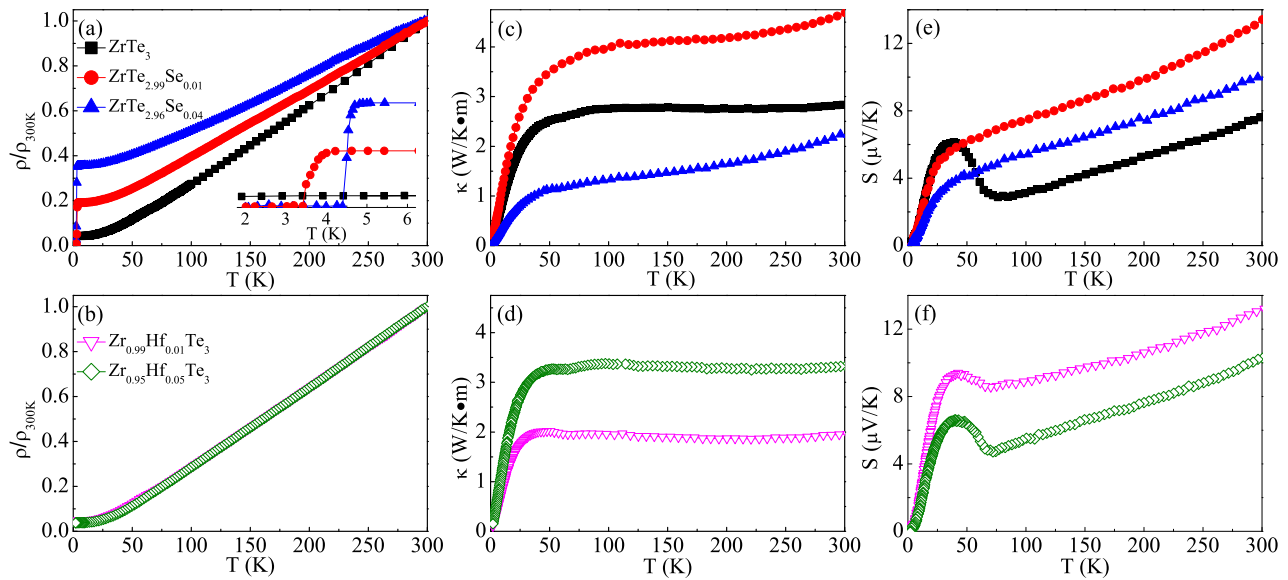}}
\caption{(Color online) Temperature dependence of (a,b) normalized resistivity $\rho$(T)/$\rho$(300 K), (c,d) thermopower $S$(T), and (e,f) thermal conductivity $\kappa$(T) for ZrTe$_{3-x}$Se$_x$ ($x$ = 0, 0.01, 0.04) and Zr$_{1-y}$Hf$_y$Te$_3$ ($y$ = 0.01 and 0.05) single crystals, respectively. Inset in (a) shows the low-temperature superconducting transitions.}
\end{figure*}

Figure 1(b) shows the normalized Zr K-edge XANES spectra. The Zr K-edge absorption energy at $\sim$ 18.008 keV for ZrTe$_3$ indicates dominant Zr$^{4+}$ state;\cite{Zahir} yet it slightly shifts to lower energies with Hf and Se doping implying valence changes. Figure 1(c) depicts the corresponding Fourier transform EXAFS of indicated samples. EXAFS could be described in a single-scattering approximation by:\cite{Prins}
\begin{align*}
\chi(k) = \sum_i\frac{N_iS_0^2}{kR_i^2}f_i(k,R_i)e^{-\frac{2R_i}{\lambda}}e^{-2k^2\sigma_i^2}sin[2kR_i+\delta_i(k)],
\end{align*}
where $N_i$ is the number of neighbouring atoms at a distance $R_i$ from the photoabsorbing atom. $S_0^2$ is the passive electrons reduction factor, $f_i(k, R_i)$ is the backscattering amplitude, $\lambda$ is the photoelectron mean free path, $\delta_i$ is the phase shift, and $\sigma_i^2$ is the correlated Debye-Waller factor measuring the mean square relative displacement of the photoabsorber-backscatter pairs. For ZrTe$_3$, the main peak at 2.76 {\AA} corresponds to the six nearest Zr-Te bonds, while the peaks at 3.18 and 3.52 {\AA} can be assigned to the next-nearest Zr-Te1 and Zr-Zr bonds, respectively. With Hf and Se substitution, all three peaks shift to smaller bond distances of 2.67, 3.13, and 3.50 {\AA}, reflecting smaller unit cell volumes of HfTe$_3$ and ZrSe$_3$ when compared to ZrTe$_3$.\cite{Zhu,Saleem,PatelKR} This indicates increased metal-chalcogen hybridization.

Figure 2(a,b) exhibits the temperature dependence of normalized resistivity $\rho/\rho_{\textrm{300 K}}$. It shows a typical metallic behavior without resistivity anomaly for all investigated samples. An abrupt resistivity drop is seen in Se-doped ZrTe$_3$ [inset in Fig. 2(a)], signaling the onset of SC. Zero resistivity is observed at $T_\textrm{c}$ = 4.4 and 3.4 K for ZrTe$_{2.99}$Se$_{0.01}$ and ZrTe$_{2.96}$Se$_{0.04}$, respectively. In contrast to ZrTe$_{3-x}$Se$_x$, no SC above 2 K was observed for Zr$_{1-y}$Hf$_y$Te$_3$ [Fig. 2(b)]. Figure 2(c,d) shows the temperature dependence of thermal conductivity $\kappa$(T) for the indicated samples. The room temperature $\kappa$ shows a relatively low value of 2.83 W/K$\cdot$m for pure ZrTe$_3$, caused by the combination of low crystal symmetry and chemical composition with heavy elements and lower than in polycrystalline samples due to the absence of grain boundaries. A kink in $\kappa$(T) is observed at $T_{\textrm{CDW}}$ for ZrTe$_3$ and Zr$_{1-y}$Hf$_y$Te$_3$, indicating strong electron-phonon coupling, which is absent in ZrTe$_{3-x}$Se$_x$. Moreover, all the $\kappa$(T) data are weakly temperature-dependent above 100 K. The absence of a commonly observed maximum in $\kappa$(T) is probably related to its rather low value, demonstrating a significant acoustic phonon scattering.\cite{Bar}

Figure 2(e,f) displays the temperature dependence of thermopower $S$(T) for the indicated samples. All the values of $S$(T) are positive, indicating dominant hole-like character of the 3D FS sheet at the BZ center. In the high-temperature regime, the $S$(T) curve is weakly temperature-dependent and shows a quasi-$T$-linear behavior. With decreasing temperature, the $S$(T) of ZrTe$_3$ changes its slope below $T_{\textrm{CDW}}$, reflecting the reconstruction of Fermi surface, in contrast to no anomaly detected in resistivity and in agreement with previous report.\cite{Felser}

In general, the $S$(T) is discussed in terms of two contributions, i.e. the diffusion term $S_{\textrm{diff}}$, and the phonon drag contribution $S_{\textrm{drag}}$ due to electron-phonon coupling. The $S_{\textrm{drag}}$ term gives $\propto T^3$ for $T \ll \Theta_\textrm{D}$, $\propto T^{-1}$ for $T \gg \Theta_\textrm{D}$, and a peak structure at $\sim \Theta_\textrm{D}/5$, where $\Theta_\textrm{D}$ is the Debye temperature. The peak feature at 37(5) K in pure ZrTe$_3$ might be attributed from the phonon-drag effect since the peak temperature is very close to $\Theta_\textrm{D}/5 \approx$ 36.8(1) K. However, the phonon drag should diminish by $T^{-1}$ at high temperature which is not found here, pointing to the presence of diffusion contribution as well. With Se doping at Te sites, there is no CDW anomaly as well as peak feature. In contrast, the Hf substitution stabilizes the CDW order in Zr$_{1-y}$Hf$_y$Te$_3$ with the $T_{\textrm{CDW}}$ gradually shifting to 72 K for Zr$_{0.95}$Hf$_{0.05}$Te$_3$ [Fig. 2(f)].

At low temperature, the diffusive Seebeck response of Fermi liquid dominates and is also expected to be linear in $T$ [Fig. 3(a)]. In a single-band system, $S$(T) is given by
\begin{equation}
\frac{S}{T} = \pm \frac{\pi^2}{2}\frac{k_\textrm{B}}{e}\frac{1}{T_\textrm{F}} = \pm\frac{\pi^2}{3}\frac{k_\textrm{B}^2}{e}\frac{N(\varepsilon_\textrm{F})}{n},
\end{equation}
where $e$ is the electron charge, $k_\textrm{B}$ is the Boltzmann constant, $T_\textrm{F}$ is the Fermi temperature, which is related to the Fermi energy $\varepsilon_\textrm{F}$ and the density of states $N(\varepsilon_\textrm{F})$ as $N(\varepsilon_\textrm{F})$ = $3n/2\varepsilon_\textrm{F}$ = $3n/k_\textrm{B}T_\textrm{F}$, and $n$ is the carrier concentration (the positive sign is for hole and the negative sign is for electron).\cite{Barnard,Miyake} In a multiband system, it gives the upper limit of the $T_\textrm{F}$ of the dominant band. The derived value of $S/T$ from 5 to 20 K is $\sim$ 0.292(3) $\mu$V/K$^2$ for ZrTe$_3$. It changes to $\sim$ 0.252(5) and 0.156(3) $\mu$V/K$^2$ for ZrTe$_{2.99}$Se$_{0.01}$ and ZrTe$_{2.96}$Se$_{0.04}$, respectively. We obtain the $T_\textrm{F}$ $\sim$ $1.69(3)\times10^3$ and $2.72(5)\times10^3$ K for ZrTe$_{2.99}$Se$_{0.01}$ and ZrTe$_{2.96}$Se$_{0.04}$, respectively. The ratio of $T_\textrm{c}/T_\textrm{F}$ characterizes the correlation strength in bulk superconductors. For instance, $T_\textrm{c}/T_\textrm{F}$ is close to 0.1 in Fe$_{1+y}$Te$_{1-x}$Se$_x$, pointing to the importance of electronic correlation;\cite{Pourret} while it is $\sim$ 0.02 in BCS superconductor LuNi$_2$B$_2$C. The value of $T_\textrm{c}/T_\textrm{F}$ is 0.0020(1) for ZrTe$_{2.99}$Se$_{0.01}$ and 0.0016(1) for ZrTe$_{2.96}$Se$_{0.04}$, respectively. Hence, electronic correlation probed by the $b$-axis thermal transport in ZrTe$_{2.96}$Se$_{0.04}$ is not strong and is highly sensitive to volume changes since unit cell contraction typically bring band broadening.\cite{YonezawaF,SinghDJ} On the other hand, this points to highly anisotropic correlations since the Kadowaki-Woods scaling $A_a$/$\gamma$$^2$ is comparable to Sr$_2$RuO$_4$.\cite{XD2}

Temperature dependence of specific heat $C_\textrm{p}$(T) in zero field for ZrTe$_{2.96}$Se$_{0.04}$ is depicted in Fig. 3(b). After subtraction of the phonon part by a polynomial fit, the electronic term $C_e$(T) of ZrTe$_{2.96}$Se$_{0.04}$ shows a clear jump at $T_\textrm{c}$ [inset in Fig. 3(b)], in agreement with the resistivity and thermopower data. The observed C/T increases below 10 K and just above $T_c$ takes the value of $\gamma$ $\sim$ 3.4 mJ/mol$\cdot$K$^2$ for ZrTe$_{2.96}$Se$_{0.04}$. This is larger than those of bulk superconducting ZrTe$_3$ and (Ni,Cu)$_{0.05}$ZrTe$_3$ (2.1 $\sim$ 2.7 mJ/mol$\cdot$K$^2$).\cite{HC,XD1,Zhu} The electronic specific heat jump at $T_c$, i.e. $\Delta$C$_{e}$/$\gamma T_c\approx$ 0.22, is smaller than the weak coupling value of 1.43 for the electron-phonon mediated BCS superconductors.\cite{McMillan} As we know, the electronic specific heat can also be expressed as:
\begin{equation}
\gamma = \frac{\pi^2}{2}k_\textrm{B}\frac{n}{T_\textrm{F}} = \frac{\pi^2}{3}k_\textrm{B}^2N(\varepsilon_\textrm{F}).
\end{equation}
Combining equations (1) and (2) yields: $S/T = \pm \gamma/ne$, where the units are V/K for $S$, J/K$^2$$\cdot$m$^3$ for $\gamma$, and m$^{-3}$ for $n$, respectively. This relation was shown to hold in the $T$ = 0 limit for a lot of materials, including heavy fermion metals, organic conductors, and cuprates.\cite{Behnia} Then we can estimate a dimensionless quantity
\begin{equation}
q=\frac{S}{T}\frac{N_\textrm{A}e}{\gamma},
\end{equation}
where $N_\textrm{A}$ is the Avogadro number. The value of $q$ gives the number of carriers per formula unit, which is $\sim$ 4.4.1(1) for ZrTe$_{2.96}$Se$_{0.04}$ and the estimated carrier density per volume $n \approx 9.8(1) \times 10^{20}$ cm$^{-3}$. Then the Fermi momentum $k_\textrm{F} = (3\pi^2n)^{1/3} \approx 3.07(1)$ nm$^{-1}$, and the effective mass $m^*$, derived from $k_\textrm{B}T_\textrm{F} = \hbar^2 k_\textrm{F}^2/2m^*$, is 1.53(1) $m_e$ for ZrTe$_{2.96}$Se$_{0.04}$.

\begin{figure}
\centerline{\includegraphics[scale=1]{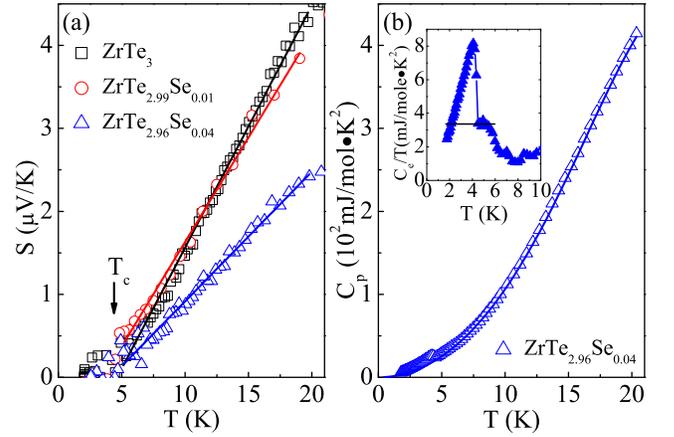}}
\caption{(Color online) (a) The low-temperature thermopower $S$(T) for ZrTe$_{3-x}$Se$_x$ ($x$ = 0, 0.01, 0.04) with linear fits from 5 to 20 K. (b) Temperature dependence of specific heat $C_\textrm{p}$(T) in zero magnetic field for ZrTe$_{2.96}$Se$_{0.04}$. Inset in (b) shows the electronic part $C/T$ after substraction of the lattice part by $C_\textrm{L}$ = $\gamma$T+$B_3T^3$+$B_5T^5$.}
\end{figure}

\begin{figure*}
\centerline{\includegraphics[scale=1]{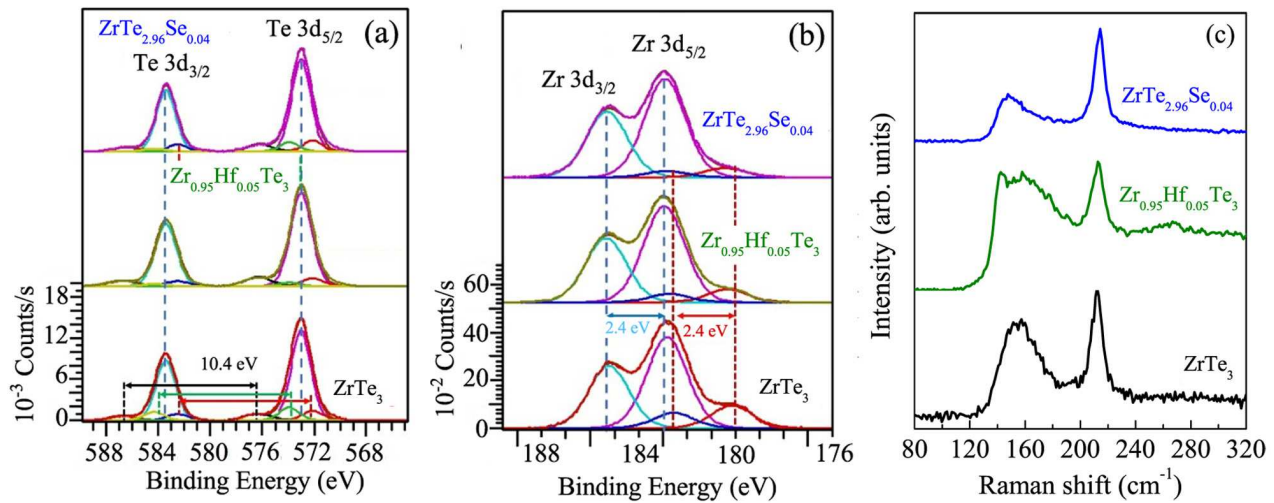}}
\caption{(Color online)(a) Te 3$d$ and (b) Zr 3$d$ XPS spectra obtained from ZrTe$_3$, ZrTe$_{2.96}$Se$_{0.04}$ and Zr$_{0.95}$Hf$_{0.05}$Te$_3$ crystals. (c) Raman peaks of the indicated crystals.}
\end{figure*}

Figure 4(a,b) shows the XPS spectra of ZrTe$_3$, Zr$_{0.95}$Hf$_{0.05}$Te$_3$ and ZrTe$_{2.96}$Se$_{0.04}$. For the analysis of Te 3$d$ peaks [Fig. 4(a)], doublets with the spin-orbit splitting separation ($\approx$ 10.4 eV), which are located at 573.0 eV (3$d$$_{5/2}$) and 583.4 eV (3$d$$_{3/2}$), 572.1 eV (3$d$$_{5/2}$) and 582.5 eV (3$d$$_{3/2}$), and 573.9 eV (3$d$$_{5/2}$) and 584.2 eV (3$d$$_{5/2}$) were deconvoluted and assigned to the Zr-Te bonds (Te atoms at the corners), Zr-Te bonds (Te atoms in neighboring chains) and Te-Te bonds after the background subtraction, respectively.\cite{YuX} We also observe peaks at 576.3 eV (3$d$$_{5/2}$) and 586.7 eV  (3$d$$_{3/2}$) that are contributed by Te-O band from the small concentration of residual oxidized Te at the surface.\cite{TohRJ,YangL} Zr-Te bonds at the ZrTe$_6$ prism corners are contributed by all Te atoms, i.e. by Te1, Te2 and Te3 atomic positions whereas Te-Te bonds in chains are contributed by Te2 and Te3 atoms only [Fig. 1(a)]. In the Zr 3$d$ spectra [Fig. 4(b)], two sets of doublets are observed with spin-orbit splitting separation ($\approx$ 2.4 eV) located at 180 eV (3$d$$_{5/2}$) and 182.4 eV (3$d$$_{3/2}$), and 183.0 eV (3$d$$_{5/2}$) and 185.4 eV (3$d$$_{3/2}$) after the background subtraction, respectively. Such observation is quite similar to the XPS analysis of ZrTe$_3$ and HfTe$_3$ in which mixed-valent states of metal atoms were assumed.\cite{JingLi,YuX} Therefore, we also consider that Zr$^{4+}$ (doublets located at the high-energy part) and Zr$^{2+}$ (doublets located at the low-energy part) might coexist. The atomic concentration ratio of Zr$^{2+}$/Zr$^{4+}$ is lower in ZrTe$_{2.96}$Se$_{0.04}$ (8.8\%) when compared to Zr$_{0.95}$Hf$_{0.05}$Te$_{3}$ (13.5\%) and ZrTe$_3$ (25.1\%). The positions of characteristics Raman peaks for all investigated crystals around 215 and 155 cm$^{-1}$ are shown in [Fig. 4(c)]. Peak positions, the relative peak intensity and the absence of Raman peaks at lower wave numbers for pure ZrTe$_3$ are consistent with previous observations on freshly exfoliated ZrTe$_3$.\cite{YangS} The wide Raman peak at lower wave numbers is wider in Zr$_{0.95}$Hf$_{0.05}$Te$_{3}$. In addition, we observe a small red shift in ZrTe$_{2.96}$Se$_{0.04}$ to lower wave numbers ($\sim$ 145 cm$^{-1}$). Both observed peaks correspond to $A_g$ phonon modes which involve atomic movements in the $ac$ plane.\cite{Hu} The wide Raman mode at low wave numbers is strongly coupled to continuum of electronic excitation and is a fingerprint of electron-phonon interaction, involving longitudinal deformation of Te2-Te3 chains in its eigenvector.\cite{Hu}

When compared to ZrTe$_3$ [Fig. 4(a)], the relative intensity of Te2-Te3 bonds at 573.9 eV is somewhat decreased for Se-doped sample. The reduced intensity of Te2-Te3 bonds at 573.9 eV could indicate some Te2-Te3 bonds breaking, or the Te2-Te3 bonds relaxing that leads Te 3$d$ binding energy decrease to energy range around 573 eV where it can not be distinguished with the major Te-Zr bond at the prism corner. In either case, a small part of the Te2-Te3 lattice chain in the $a$-axis direction is randomly relaxed or distorted due to random substitution of dopant Se atoms. Since the CDW originates in the Fermi surface pocket of the Te2-Te3 orbital character, CDW suppression mechanism by Se doping is associated with randomly disordered Te2-Te3 lattice chain in $a$-axis direction. Due to Te2-Te3 bond relaxation, the Raman active Ag mode vibration in the $ac$ plane becomes easier, which is consistent with the 10 cm$^{-1}$ Raman red shift at 215 cm$^{-1}$ [Fig. 4(e)].

When compared to ZrTe$_3$, in Hf-doped crystal Te2-Te3 lattice chain in $a$-axis direction, most Te2-Te3 bonds are relaxed as the intensity of Te2-Te3 bonds almost vanished. In addition, the relative intensity of Zr-Te bonds (Te in neighboring chains) is somewhat reduced due to the substitution of Zr by Hf. The entire Te2-Te3 chain relaxation does not affect its periodicity, hence it preserves CDW. Moreover, since longitudinal deformations of the Te2-Te3 chain exhibit strong interactions with the conduction electrons,\cite{Hu} the electron-phonon coupling becomes easier, consistent with the rise in $T_{\textrm{CDW}}$. Though the longitudinal deformations of Te2-Te3 chain also increase electron-phonon coupling in Se-doped sample, random disorder in Te2-Te3 lattice affects the nesting condition.\cite{HoeschM2} Wider Raman peak at low wave numbers is likely due to disorder at the Zr atomic position with Hf substitution. Though the Te-Zr bonds at the prism corners are same as their Te 3$d$ binding energy are unchanged at 573 eV [Fig. 4(a)], the spatial distribution of Te around Zr is not spatially uniform. This creates an unequal charge orbital environment and consequently inhomogeneous Zr valence.

In summary, our study shows that electronic correlations in superconducting ZrTe$_{2.96}$Se$_{0.04}$ viewed by thermal transport along the $b$-axis of the unit cell are not strong and are governed by the unit cell contraction as more Se is added in the lattice. Hf substitution on Zr site in ZrTe$_3$ increases $T_{\textrm{CDW}}$ whereas it is rapidly suppressed with Se substitution on Te site due to disorder in Te2-Te3 atomic chains whose orbitals form band with nesting condition at the Fermi surface. The mixed valence disparity of Zr in ZrTe$_3$, Zr$^{2+}$ and Zr$^{4+}$ is reduced in doped crystals but superconductivity with large increase in $T_c$ emerges only if CDW is suppressed. Since even modest correlations induce considerable scattering anisotropy in ZrTe$_{2.96}$Se$_{0.04}$,\cite{XD2} and since ZrTe$_3$ is considered for interconnects in the next generation room-temperature nanoscale semiconductor technology due to its size-independent low resistivity, high breakdown current density that increases with size reduction,\cite{WenX,Geremew,Geremew1} future nanoscale devices that would exploit anisotropic properties of ZrTe$_3$-doped crystals are also of considerable interest.

Work at BNL is supported by the Office of Basic Energy Sciences, Materials Sciences and Engineering Division, U.S. Department of Energy (DOE) under Contract No. DE-SC0012704. This research used the 8-ID beamline of the NSLS II, and resources of the Center for Functional Nanomaterials (CFN), which is a U.S. Department of Energy Office of Science User Facility, at Brookhaven National Laboratory under Contract No. DE-SC0012704.

\section{AUTHOR DECLARATIONS}
\subsection{Conflict of Interest}
The authors have no conflicts to disclose.
\subsection{DATA AVAILABILITY}
The data that support the findings of this study are available from the corresponding authors upon reasonable request.\\
$^{\S}$Present address: Los Alamos National Laboratory, Los Alamos, NM 87545, USA.\\
$^{*}$Present address: Anhui Province Key Laboratory of Condensed Matter Physics at Extreme Conditions, High Magnetic Field Laboratory, Chinese Academy of Sciences, Hefei 230031, China.\\
$^{\|}$Present address: Department of Physics and Beijing Key Laboratory of Opto-electronic Functional Materials $\&$ Micro-nano Devices, Renmin University of China, Beijing 100872, China.\\
$^{\P}$Present address: ALBA Synchrotron Light Source, Cerdanyola del Valles, E-08290 Barcelona, Spain.\\
$^{\dag}$yuliu@lanl.gov
$^{\ddag}$petrovic@bnl.gov

\end{document}